# Tail Risk of Electricity Futures

Juan Ignacio Peña[a], Rosa Rodríguez[b]

, and Silvia Mayoral[c]

## Abstract

This paper compares the in-sample and out-of-sample performance of several models for computing the tail risk of one-month and one-year electricity futures contracts traded in the NordPool, French, German, and Spanish markets in 2008-2017. As measures of tail risk, we use the one-day-ahead Value-at-Risk (VaR) and the Expected Shortfall (ES). With VaR, the AR (1)-GARCH (1,1) model with Student-t distribution is the best-performing specification with 88% cases in which the Fisher test accepts the model, with a success rate of 94% in the left tail and of 81% in the right tail. The model passes the test of model adequacy in the 100% of the cases in the NordPool and German markets, but only in the 88% and 63% of the cases in the Spanish and French markets. With ES, this model passes the test of model adequacy in 100% of cases in all markets. Historical Simulation and Quantile Regression-based approaches misestimate tail risks. The right-hand tail of the returns is more difficult to model than the left-hand tail and therefore financial regulators and the administrators of futures markets should take these results into account when setting additional regulatory capital requirements and margin account regulations to short positions.

*a*. Corresponding author. Universidad Carlos III de Madrid, Department of Business Administration, c/ Madrid 126, 28903 Getafe (Madrid, Spain). ypenya@eco.uc3m.es; *b*. Universidad Carlos III de Madrid, Department of Business Administration, c/ Madrid 126, 28903 Getafe (Madrid, Spain). rosa.rodriguez@uc3m.es *c*. Department of Business Administration, c/ Madrid 126, 28903 Getafe (Madrid, Spain). *c*. Universidad Carlos III de Madrid, Department of Business Administration, c/ Madrid 126, 28903 Getafe (Madrid, Spain). silvia.mayoral@uc3m.es . Department of Business Administration, c/ Madrid 126, 28903 Getafe (Madrid, Spain). An anonymous referee provided many useful comments. We acknowledge financial support from FUNCAS, through grant PRELEC2020-2017/00085/00, from DGICYT, through grant ECO2016-77807-P, and from CAM, through grant EARLYFIN-CM, #S2015/HUM-3353.

## 1. Introduction

The liberalization of electricity markets generates competition coupled with price uncertainty. Both factors create powerful incentives to carry out effective management of the market risk of electricity prices. Consider a portfolio of an energy trading company. If it trades in gas and power, it may have positions in many locations. Each location is characterized by its forward price and volatility curves. Thus, the firm may face hundreds of portfolio value drivers[1]. Given this high dimensionality, mapping the changes in portfolio value on a reduced set of principal risk factors is common practice in the industry, Eydeland, and Wolyniec (2003). Electricity futures contracts prices are key risk factors and we need thus methods for representing their distribution for the computation of tail risk measures of complex portfolios using Monte Carlo simulation or delta-gamma approaches.

The right assessment of the tail risk of the price of electricity futures contracts is crucial when setting the size of the buffer designed to protect the futures market's clearinghouse from losses caused by defaults in traders' positions. Most futures markets erect at least two consecutive safeguards. First, the exchange's own default fund and second, the clearinghouse members' default fund. If, when setting the size of both default funds, they underestimate losses because of extreme movements in futures prices, the exchange and its members may face significant liquidity (and perhaps solvency) problems. A recent event illustrates this situation. On September 13, 2018, the Financial Times (2018), reported that in the Nasdaq's Nordic futures power market the losses incurred by just one trader[2] wiped out all the exchange's own default fund of €7 million and an additional €107 million, or 64%, of the member's default fund of €166 million. These losses prompted Nasdaq to ask the members of the clearinghouse to make compensating payments to the funds within 48 hours.

---

[1] Eydeland and Wolyniec (2003) suggest that in the US the firm may face several thousand possible value drivers.
[2] The trader, an individual member of the Nasdaq Clearing firm, could clear his own trades without going through a bank. He based his strategy on betting on a narrowing of the price differential between German and Norpool electricity prices. When implementing this strategy, he likely bought Norpool futures and sold Germany's EEX futures. German prices augmented because of increases in carbon prices and, at the same time, Nordic prices fell because of wetter than previously expected weather expanded the hydropower supply. Both legs of the strategy generated huge losses simultaneously and, unable to meet the margin calls, the trader declared bankruptcy.

Although research exists (Byström, 2005, Westgaard et al., 2019b) about what method provides better forecasts of tail risk in day-ahead electricity prices, little is known about this issue with longer-maturity contract prices (monthly, quarterly and yearly). The scarcity of research on these matters implies a lack of convincing evidence on how well alternative models measure the tail risk of the electricity futures price. In this paper, we aim to fill this gap in the literature, by comparing the in-sample and out-of-sample performance of several models and markets and suggesting policy recommendations, based on the empirical evidence provided.

The non-storability of electricity and the physical and reliability constraints in its transmission manifest itself in complex price distributions, as measured by time-dependent volatility, skewness, kurtosis, and quantiles. Although research on the distributional characteristics of spot and day-ahead electricity prices is available (e.g. Escribano et al., 2011), we know little about the tail risk of futures prices with maturities longer than one day and its implications for risk management. In this paper, we compare the in-sample and out-of-sample performance of several models for computing the tail risk of electricity futures contracts with one month and one year to maturity traded in the NordPool, French, German and Spanish markets in 2008-2017. As measures of tail risk, we use Value-at-Risk and Expected Shortfall. Our findings support the conclusion that the AR (1)-GARCH-t model performs better than other models in predicting VaR and ES in all markets. Overall, and with VaR, the AR(1)-GARCH(1,1) model with Student-t distribution is the best model with 88% cases in which their adequacy is accepted by the Fisher test, with a higher success rate in the left tail (94%) than in the right tail (81%). This model works better for monthly contracts than for yearly contracts and passes the test of model adequacy in the 100% of the cases in the NordPool, and German markets, and in 88% of the cases in the Spanish market, but only in the 63% of the cases when predicting VaR in the French market. We report persistent risk misestimation in widely used methods in the industry, such as Historical Simulation or methods that seem to work well in other energy markets, such as Quantile Regression. The right-hand tail of the returns is more difficult to model than the left-hand tail.

This study makes several contributions to existing literature. First, while several studies (e.g. Byström, 2005, Chan and Grey, 2006) have addressed the performance of VaR forecasting models in electricity prices, few papers have focused on the estimation of extreme high and low quantiles for the prices of electricity futures contracts. With this paper, we aim to fill this gap in the literature. This is important because producers and retailers of electricity pursuing active risk management policies use futures contracts of different maturities (monthly, quarterly, yearly) when they become available (Sanda et al., 2013, Boroumand et al., 2015). In contrast with Gonzalez-Pedraz et al. (2014) who studied the tail risk of monthly futures contracts in one market (NYMEX), we study monthly and yearly maturities in several markets. Second, we model the tail risk both for long positions in futures contracts, typically held by retailers and for short positions in a futures contract, typically held by producers. Third, earlier studies focused on the NordPool or German markets. We study those markets, but we also include the French and Spanish markets, which have received limited attention in the literature so far. Fourth, we study an array of competing models, from the simple historical simulation to quantile regression. Fifth, we study the extent to which the model's out-of-sample performance matches its in-sample fitting. Also, besides VaR, this paper presents the forecast and backtesting of ES, which is becoming the tail risk measure of choice for regulators.

We organize the rest of this paper as follows. Section 2 reviews the literature. After describing the methodology in Section 3, we present data in Section 4. Section 5 discusses the empirical results. Section 6 concludes.

## 2. Literature Review

Literature has addressed optimal hedging strategies in those markets using financial contracts, Deng and Oren (2006). Vehviläinen and Keppo (2003) studied the optimal hedging strategy of spot price risk by using futures contracts. Most papers investigating the tail risk of electricity prices focus on the spot or day-ahead price and study hourly prices or their daily averages. However, we cannot translate the results directly to the prices of futures contracts with longer maturities. Spot or day-ahead prices trade seven days a week and the average value and the volatility differ between hours and between weekdays and weekends. Also, these prices present strong seasonality, regime-switching, large price spikes, time-varying daily volatility (Knittel and Roberts, 2005, Escribano et al., 2011, Paraschiv et al., 2016), complex marginal distributions with significant skewness and excess kurtosis (Benth et al., 2008) and negative prices (Fanone et al., 2013). Futures prices trade five days a week and are always non-negative, the seasonal variation is less salient, price spikes are less frequent and smaller, the average price increases with maturity and the volatility (Samuelson effect), skewness and kurtosis decrease with maturity (Blanco et al., 2018), but they may present regime-switching and non-Gaussian marginal distributions (Andresen et al., 2010). As Quinn, et al. (2005) argue, futures prices are a function of market expectations of demand and cost conditions during the delivery period. The current market situation (i.e. spot prices) may or may not influence these expectations and as a result, the tail risk may be different for spot and futures prices.

Most papers investigating the tail risk in electricity futures prices focus on day-ahead prices. Henney and Keers (1999) suggest the Historical Simulation (HS) in the case of established electricity markets. Leong and Siddiqi (1999) compute the VaR of electricity spot prices assuming a normal distribution for electricity prices. Dahlgren et al. (2003) apply the historical simulation method to the computation of VaR and ES of day-ahead electricity prices. Byström (2005) studies hourly day-ahead Nord Pool prices using a GARCH-EVT framework and recommends the generalized Pareto distribution. Liu and Wu (2007) use the delta-normal method to calculate the value of VaR of day-ahead electricity prices. Keles et al. (2016) report similar results to Byström, (2005) in the case of the German market (EPEX). Chan and Grey (2006) analyze day-ahead prices from five markets using and EGARCH-EVT specification that performs well in forecasting out-of-sample VaR in comparison to several models. Bunn et al. (2016) study UK day-ahead prices and conclude that a linear quantile regression model outperforms skewed GARCH-t and CAViaR models in out-of-sample forecasts of VaR. Paraschiv et al. (2014, 2016) show the improvements got in forecasting the tail risk of day-ahead prices by using fundamental variables, and Hagfors et al. (2016a) using quantile regression techniques show that the sensitivities to fundamentals may depend on market conditions. Hagfors et al. (2016b) analyze extreme positive and negative price occurrences in the German day-ahead market by using non-linear methods. Westgaard et al. (2019b) present a detailed analysis of the performance of VaR forecasts of hourly day-ahead prices in the German market, using fundamental explanatory variables and six different models: QR, EWQR, EWDKQR, GARCH-T, SAV CAViaR and AS CAViaR. Although the best-performing model varies across trading periods and depending on the evaluation criteria, the authors recommend the EWQR based on selected fundamentals as the best model overall.

With futures with longer maturities, as far as we know, the evidence on the relative performance of tail risk measures is scarce. Gonzalez Pedraz et al. (2014) analyze PJM electricity futures with monthly maturity (M1, the front contract) traded in the NYMEX market in an energy portfolio analysis and report that the best VaR forecasting model is a GARCH model with a Student t distribution. Westgaard et al. (2019a) investigate the in-sample performance of estimates of one-day-ahead VaR got from three univariate models (EWMA, FHS, QR) of the prices of nine European energy futures front (M1) contracts, for oil, gas, coal, and M1, M2 and M3 contracts for electricity. They conclude that VaR forecasts made by the QR model present the best conditional and unconditional performance, followed by the FHS, whereas EWMA's forecast is the poorest. Also, the modeling of the right-hand tail is more challenging in all cases, suggesting that the tail risk attributable to short positions is more difficult to assess.

3. Methodology

This section presents the various approaches to calculating VaR($\alpha$,t) and ES($\alpha$,t) for quantile $\alpha$ and period $t$ examined in this paper. These distribution-based risk measures are estimates of the capital reserve (as required by the regulator or set by the firm) as protection against fluctuations in the value of a financial asset. We focus on one-step-ahead estimates of VaR($\alpha$,t) and ES($\alpha$,t) calculated with data up to period $t$ -1 and based on daily returns for long ($\alpha$ = 0.05, 0.01) and short ($\alpha$ = 0.95, 0.99) positions. Let $r_j$ be the asset return series where $j = t$-1,…,1 . We define the left-hand side VaR($\alpha$,$t$), measuring the tail risk of long futures positions, as

$$\Pr[r(t) < VaR(\alpha, t)] = \alpha \ ; \ \alpha < 0.5 \qquad (1)$$

The right-hand side VaR($\alpha$,t), measuring the tail risk of short futures positions, is

$$\Pr[r(t) > VaR(\alpha, t)] = 1 - \alpha \ ; \ \alpha \geq 0.5 \qquad (2)$$

In the same way, we define the left-hand side ES($\alpha$,t) as

$$ES(\alpha, t) = E[r(t)|\ r(t) < VaR(\alpha, t)] \ ; \ \alpha < 0.5 \qquad (3)$$

And the right-hand side ES($\alpha$,t) as

$$ES(\alpha, t) = E[r(t)|\ r(t) > VaR(\alpha, t)] \ ; \ \alpha \geq 0.5 \qquad (4)$$

Regarding the computation of ES($\alpha$,t), we apply two methods. First, if the model we use allows a closed-form for its computation, we use it. Second, if there is no closed-form available, we rely on the representation of ES as integrated VaR as presented in Acerbi and Tasche (2002)

$$ES(\alpha, t) = \frac{1}{\alpha} \int_0^\alpha VaR(u) du \qquad (5)$$

and apply the discretization suggested in Emmer et al. (2015)

$$ES(\alpha, t) \cong \frac{1}{4}(VaR(\alpha, t) + VaR(0.75\alpha + 0.25I(\alpha), t) + VaR(0.5\alpha + 0.5I(\alpha), t)$$

$$+ VaR(0.25\alpha + 0.75I(\alpha), t) \qquad I(\alpha) = \begin{cases} 1 & \alpha \geq 0.5 \\ 0 & \alpha < 0.5 \end{cases} \qquad (6)$$

Intuitively (6) suggests that an estimate of ES could be considered reliable if estimates of the four VaR are reliable[3]. Backtesting ES could be done by simultaneously backtesting multiple VaR estimates at different levels.

As methods for the computation of those risk measures, we apply two non-parametric models. First, we consider the Historical Simulation (HS) method because of Perignon and Smith (2010) report that one-day-ahead VaR(0.01) based on historical simulation is the standard in the case of international commercial banks. Second, we apply the Boudoukh-Richardson-Whitelaw (Boudoukh et al., 1998) method of exponentially declining weights of historical returns because Taylor (2008) commends this method as an equivalent to Exponentially Weighted Quantile Regression (EWQR) with one intercept and no regressors and shows that they may be able to outperform GARCH-based methods and CAViaR models. We also apply parametric approaches based on AR(1)-GARCH(1,1), AR(1)-EGARCH(1,1) or AR(1)-GJR(1,1) models with Student t distributions which are considered as standard benchmarks for the computation of risk measures (e.g. Du and Escanciano, 2017). Also, we apply quantile regression (QR) models which have been used successfully in modeling extreme quantiles in electricity markets by Hagfors et al. (2016a,b) among others. Table 1A presents the models.

[INSERT TABLE 1A HERE]

---

[3] For instance, ES(0.05) should be based on estimates of VaR(0.05), VaR(0.0375), VaR(0.025) and VaR(0.0125)

Next, we present the backtesting methods for VaR (Bin, POF, CCI, and DQ) and ES (UN-N, and UN-t) [4]. The following Table 1B summarizes their principal characteristics

[INSERT TABLE 1B HERE]

With VaR, two tests (Bin, POF) focus on the proportion of failures (unconditional failure) and two tests on conditional failure for clustering (CCI, DQ). Besides results from traditional individual tests (e.g., POF, CCI), and to give an overall perspective of the results from the VaR backtesting exercise we present a combined p-value test for multiple hypothesis testing. This procedure follows the same principle of the mixed coverage test, Haas (2001). We use Fisher's product test combining p-values to test the global null hypothesis $H_0$ that each of n component null hypotheses, $H_1, \ldots, H_n$ is true versus the alternative that at least one of $H_1,\ldots, H_n$ is false. As Zhang et al. (2013) show, Fisher's test is more powerful than classical multiple tests such as the Bonferroni test and the Simes tests, especially when n is not large. The combined p-value test is based on the notion that several non-significant results together may suggest significance and hence detect departures from $H_0$. The test is

$$F = -2 \sum_{i=1}^{n} \ln(p_i) \qquad (7)$$

that follows a $\chi^2_{2n}$ distribution under $H_0$ and the assumption that the tests are independent. In the empirical study, we assume that the test (7) rejects the null hypotheses of model adequacy when its p-value is below 1%. The tables show the number of times and percentages that data do not reject the overall Fisher test at 1% level for VaR and ES models, contracts, and countries.

---
[4] The technical details on models and backtesting test are in Appendix A.

With ES, backtesting is more difficult than with VaR, Hull, and White (2014). ES estimates rely on a smaller number of observations than VaR estimates and are thus less stable, especially when the return distribution has fat tails (Yamai and Yoshiba, 2002). For instance, in the current Basel III regulatory framework, ES estimators are based on only 2.5% of the sample size used for estimating the VaR[5]. With ES, both unconditional backtests (UN-N and UN-t) compare observed and expected ES, but assuming different reference distributions (Normal and Student-t). The expected value for the UN test statistic is zero, and it is negative when there is evidence of risk underestimation. To determine how negative it should be to reject the model, we need critical values, based on distributional assumptions (Normal, Student-t) for the price returns. We check the consistency of the tests' outcomes and report the p-value of each UN test separately. Notice that consistency implies that (i) if the UN-N test shows no undervaluation, the UN-t test must match, (ii) if the UN-t test suggests undervaluation, the UN-N test must match. An inconsistency arises when the UN-N test shows no undervaluation but the UN-t test says the opposite.

**Data Availability**

Datasets related to this article can be found at EEX, OMIP, and NASDAQ OMX, hosted at http://www.eex.com, www.omipo.pt, and http://www.nasdaqomx.com/

4. **Data**

---

[5] Standard back-testing periods of 250 days are routinely recommended by regulators. Using the 2.5% confidence limit gives 250*0.025= 6.25 failures.

We collect daily prices of baseload futures contracts with one-month and one-year maturity in the period from January 2, 2008, to September 25, 2017, from the German, French, Spanish and NordPool markets. Price time series are continuous and computed as perpetually linked series[6]. For instance, in the case of the month series, the first data point on January 2, 2008, corresponds to the price of the front contract M1 to supply electricity during February 2008. On February 1, 2008, the price in the continuous series corresponds to the price of the front contract M1 to supply electricity during March 2008, and so on. Table 1C has a contract definition, codes, and sample period of all contracts and markets.

[INSERT TABLE 1C HERE]

In what follows, we consider the full sample period[7] (January 2, 2008, to September 25, 2017), the in-sample period for model estimation (January 2, 2008, to December 31, 2014) and the out-of-sample period for backtesting purposes (January 2, 2015, to September 25, 2017). Figure 1 displays the prices of the futures contracts in the full sample, Panel A contains M1 contracts and Panel B shows Y1 contracts.

[INSERT FIGURE 1 HERE]

After steep increases during 2009, the prices of the month contract show a downward trend in the German market and, to a lesser extent, in the French and NordPool markets, but are steady in the Spanish market. Noticeable price increases, including price spikes, are apparent in the French market during November 2016 because of the combination of low temperatures and shortages in nuclear production because of extensive revisions in many facilities during this period (Commission de Régulation de L'Énergie, 2017). Prices of year contracts follow similar patterns to month contracts but are less volatile and spiky. The behavior across markets is alike, peaking around mid-2008, and then rapidly decreasing in all markets, bottoming near the end of 2009, in coincidence with the financial crisis 2007-2009. Then a stable period ensues until 2013 where a downward trend manifests in the French, German, and NordPool markets, bottoming out at the end of 2015. Prices in the Spanish market present a steadier pattern.

---

[6] Month contracts cease trading at the close of business, one Business Day prior to the last calendar day of the contract's delivery period. However, usually there is hardly any trading volume after the day before the first calendar day of the delivery month. Year contracts cease trading at the close of business, two Business Days prior to the first calendar day of the delivery year.

[7] The number of data points in each sample varies across markets, due to national holidays and other market-specific events. For instance, in Germany (NordPool) the figures for the full-sample, in-sample, and out-of-sample are 2470 (2438), 1772 (1756), and 698 (688) respectively.

We compute daily returns as the first difference in log prices[8]. Figure 2 describes the log returns over the full sample period and Panel A contains M1 contracts and Panel B shows Y1 contracts.

[INSERT FIGURE 2 HERE]

The graphs of the return series suggest time-varying volatility and volatility clustering around supply and demand shocks, for instance, in the French market during November 2016 or in the Spanish market during January 2014. Several positive and negative spikes are apparent, these features being more extreme in the case of the month series. Notice that a positive (negative) spike in the return series is caused by a positive (negative) level change in the price series. The literature usually relates the reasons justifying these jumps to events in the industry deemed to have a lasting impact on prices. For instance, on March 13, 2011, Angela Merkel, the German chancellor, ordered an emergency safety check at all Germany's 17 nuclear power stations in the wake of the Japanese earthquake (Financial Times, 2011a), and on March 15, 2011 (Financial Times, 2011b) she announced a decree to make seven of the country's 17 nuclear plants idle. The German M1 and Y1 returns series show positive spikes on March 15, 2011, of 15% and 6% respectively, suggesting an expectation of permanent increases in electricity prices during the corresponding delivery periods because of lower supply. Interestingly, in the French market, the month contract also shows a spike of 12% on the same date, but the year contract does not present an extreme return. Jumps are also apparent in the French market during November 2016, corresponding to the contract with delivery during December 2016 and related to lasting decreases in supply, as discussed above. In the Spanish market during December 2013 and January 2014 some positive and negative jumps bunch together, due to reduced supply in four out of the eight nuclear plants available, followed by unprecedented increases in Eolic production (CNMC, 2014).

---

[8] In the continuous price series, a level change may appear on the days when trading activity moves from the old contract to the fresh one. This "rolling" event sometimes brings forth a jump in the return series, generating high volatility. Therefore, to avoid this artificial consequence, in days with "rolling" effects, we set the returns on these days equal to zero, as is usual in the literature (e.g. Westgaard et al., 2019a, Blanco et al., 2018). Notice that those jumps are spurious outcomes induced by the definition of the continuous series, but they are void of economic consequences because the trader does not receive the return between old contracts and fresh ones. Specifically, with month contracts, we assume traders close their positions in the old contract the last Business day before the delivery period, and they take positions in the new contract on the first day of the delivery period. Therefore, we set the return on that day equal to zero. With year contracts, and following the same logic, we set the return on the first day of the delivery period equal to zero.

Table 2 provides descriptive statistics of the returns in all markets and contracts for the full sample in Panel A, the in-sample in Panel B, and the out-of-sample in Panel C.

[INSERT TABLE 2 HERE]

In Panel A, we may observe that the average return is small and negative in all cases,

suggesting a persistent decrease in futures prices. Month contracts present higher price volatility than year contracts, as expected. Daily return volatility in month contracts is highest in NordPool (2.9%), followed by the French (2.6%), Spanish (1.7%), and German (1.6%) markets, and with year contracts results are alike. Although returns are negatively skewed in the French and Spanish markets, and positively skewed in Germany and the NordPool, all of them are leptokurtic with the Spanish month contract having the largest coefficient (18.5) and the NordPool year contract having the smaller (7.18). All the returns series reject the null hypotheses of the Jarque-Bera Normality test. Tail risk in month contracts as measured by the 1% and 99% quantiles is more extreme in the NordPool (-0.079, 0.080) and French (-0.063, 0.082) markets than in the German (-0.045, 0.044) and Spanish (-0.050, 0.042) markets. The risk varies between the right and left tails, for instance, the French month contract presents a higher (0.082) right tail risk than left tail risk (-0.063). Year contracts present a lower tail risk than month contracts, as expected. The Ljung-Box test with 10 and 20 lags suggests auto-correlated returns in all cases. Results in the in-sample and out-of-sample periods are similar. The overall picture indicates that the market generating more extreme returns is NordPool, followed closely by the French market, whereas the German and Spanish markets exhibit fewer extreme returns.

We now summarize how the return distribution changes over time. We have split our data set by years. Detailed results are in Table 1A in the Appendix. The key takeaway from Table 1A is that the shape of the distribution of the return changes over time in all contracts and markets, suggesting a time-dependent tail risk. Figure 3 shows the time evolution of the volatility, skewness, and kurtosis of the month and year contracts, usually the latter presenting fewer extreme values of these statistics than the former.

[INSERT FIGURE 3 HERE]

With month contracts, we may see the volatility changing over time and reaching a peak in 2014 in Spain, in 2015 in NordPool and in 2016 in France and Germany. These peaks are possibly associated with shortages in nuclear production in Spain, Germany, and France during those years, as discussed above. Regarding skewness and kurtosis, the events of 2011 in the German market are associated with peaks in skewness and kurtosis in Germany and France, suggesting an overall increase in the risk of the month futures prices in these markets during this period. Asymmetry changes sign over time in all markets and kurtosis peaked in 2008 in Spain, in 2011 in Germany, 2014 in France, and 2015 in NordPool. All these changes point to changes also in tail risk over time.

## 5. Results and discussion

In this section, we present the results of the models used to generate VaR and ES forecasts and study their performance as measured by the backtesting tests[9]. A summary of the results is in Table 3, panels A1(VaR) and B1(ES) for all contracts and markets together, A2 and B2 for VaR with M1 and Y1 contracts, panels C2 and D2 for ES, also for M1 and Y1 contracts, and panels E1 and E2 for VaR and ES by markets, respectively[10]. Table 3 shows the number of times and percentages that data do not reject the overall Fisher test of model adequacy for all models, contracts, and markets at the 1% significance level. GT is the total for both tails. GTL is the total for the left tail (5%, 1%). GTR is the total for the right tail (95%, 99%).

[INSERT TABLE 3 HERE]

---

[9] Models' estimated parameters and residual analysis are in the Appendix A, Tables 1 and 2.
[10] Detailed results on results of individual test are in Appendix A.

With VaR (Panel A1), the best model is GARCH with 88% cases in which their adequacy is accepted by the Fisher test in both tails, with a higher success rate in the left tail (94%) than in the right tail (81%). The worst model is the HS with an acceptance rate of 22%, also showing a differential success rate between the left (31%) and the right (13%) tail. Regarding the other models, the ranking is EGARCH, GJR, EWQR, and QR with overall success rates of 81%, 75%, 44%, and 28% respectively, although the success rate is higher in the left tail than in the right tail. With ES (Panel B1), the best models are three GARCH-type models with an acceptance rate of 100% in both tails. The worst model is QR-COM, accepted in 33% of cases whereas HS and EWQR present a success rate of 63%. By contracts, Panel A2 presents VaR results for M1 and Panel B2 for Y1. GARCH and EGARCH models do equally well in both contracts, with a success rate of 88% and 81% respectively, but other models do better for M1 than for Y1. For instance, EWQR has a success rate of 56% in M1 but 31% in Y1. Thus, although the volatility of the returns of Y1 contracts is lower than the ones of M1 returns, the Y1 contract's VaR is more difficult to forecast. Panel C2 and D2 present ES results for M1 and Y1, respectively. GARCH-type models have a 100% success rate in both cases. However, other models present better performance with M1 than with Y1. For instance, EWQR and HS do pretty well with M1 (100% and 88%) but less so with Y1 (25% and 38%). Panels E1 and E2 present aggregate results by markets. With VaR, GARCH, EGARCH, and GJR have a success rate of 100% in the NordPool. In Germany, GARCH, and EGARCH present a rate of 100%. However, in the and Spanish and French markets, the success rate of the best GARCH model is 88% and 63% respectively. The performance of all models in the French market is worse than in the other three markets, excepting HS, that is rejected in all cases in the NordPool market. With ES the GARCH-type models are not rejected in 100% of cases in all markets.

An important point is whether the model's rejection arises because of risk underestimation or overestimation. In Table 4, panels A-E we compare the expected number of VaR failures with actual failures in every market and contract (eight cases). We compute the deviation defined as (Observed Failures – Expected Failures)/Expected Failures. A positive (negative) figure shows that the model underestimates (overestimates) tail risk.

[INSERT TABLE 4 HERE]

The summary results in Panel E indicates that the HS method underestimates risk, with an average deviation of 34.56% (M1) and 62.24% (Y1). The EWQR method also underestimates the risk with an average deviation of 25.64% (M1) and 37.10% (Y1). The QR-COM, underestimate risk with average deviations of 78.81% (M1) and 148.44% (Y1). The GARCH and EGARCH models misestimate risk between -0.25% and 6.54% whereas GJR overestimates risk by -2.99% (M1) and -10.83% (Y1). Risk misestimation in most cases is stronger for Y1 contracts than for M1 contracts, which agrees with similar results in Table 3.

In summary, the GARCH model presents the leading forecasting performance among the models considered, followed closely by EGARCH and then by GJR while the EWQR, HS, and QR-COM models are least accurate.

As an illustration of the reasons why GARCH-type models outperform the other models, the following Figure 4 presents the VaR(1%) forecast in the out-of-sample (o-o-s) period with all models for the M1 contract in the German market. The results for ES(1%) are very similar and available on request.

[INSERT FIGURE 4]

The graph in the upper left-hand side labeled "Backtesting GARCHs" shows the results for AR(1)-GARCH(1,1), AR(1)-EGARCH(1,1), and AR(1)-GJR(1,1). The shape of the forecast is similar across the three models and follows closely the temporal evolution of the returns. When a relatively large return appears, the VaR forecast moves up (in absolute terms) after a one-day lag and then goes back to normal levels quickly. If a cluster of extreme returns appears, the forecast adjusts to that accordingly. This fast adaptability is an embedded feature of GARCH models and one of the reasons for their popularity. In contrast, HS models produce a constant VaR forecast, that changes only when very large returns appear. After that, the VaR forecast remains at the new level for the number of days (e.g. 250) in that the rolling window contains that extreme return, even when consecutive returns are small. Therefore, the HS method tends, at first, to underestimate risk and, after one big return, to overestimate risk. Given that HS only relies on the recent (one year) unconditional distribution of the risk factors, is likely to misestimate the changes in conditional risk (Pritsker, 2006). Besides, the VaR forecast generated by the HS model is largely insensitive to volatility clustering if the returns in the cluster are not very high. Similar behavior to HS is produced by the EWQR, although this model presents more adaptability to volatility clustering. Finally, the Quantile Regression (QR-COM) model generates almost constant VaR forecasts because the regressors have very low explanatory power, and therefore hardly impact the VaR forecast. Table 3 in the Appendix shows the correlations between the returns and the explanatory variables. They range from -0.12 to 0.05 and are statistically not different from zero.

The distribution of electricity futures returns exhibit heavier-than-normal tails. The tails of the return distribution decay in a power-law speed instead of the exponential speed as in that of the normal distribution. The GARCH-t model captures volatility clustering by modeling the dynamic of volatility, leading to a time-dependent estimate of the VaR and ES. Besides, the GARCH-type models capture the heavy tails, see Davis and Mikosch (2009), because GARCH processes have power-law marginal tails. In this paper, we tested the GARCH-t (also EGARCH and GJR) model instead of the most often applied GARCH-Normal because the evidence in McNeil and Frey (2000) show that the GARCH model with Student-t innovations is more efficient in forecasting the VaR, whereas the estimates from GARCH models with normal innovation underestimate tail risk (see also Mikosch and Starica, 2000). Next, we present the calculation of the tail index under the GARCH(1,1) model following Davis and Mikosch (2009). Let $X_t$ be the (filtered of AR (1) terms) return series defined as

$$X_t = \sigma_t Z_t \tag{8}$$

Where

$$\sigma_t^2 = \alpha_0 + \alpha_1 X_{t-1}^2 + \beta_1 \sigma_{t-1}^2 \tag{9}$$

Where $Z_t$ is an iid sequence with mean zero and unit variance and the process (9) is second-order stationary if $(\alpha_1 + \beta_1) < 1$, Nelson (1990). The tails of the marginal distribution of a stationary GARCH(1,1) process have a power-law shape. This implies that there is a tail index $k > 0$ and a positive constant $c$ such that

$$P(|X| > x) \sim c x^{-2k} \tag{10}$$

The tail index *k* solves the equation

$$E[(\alpha_1 Z^2 + \beta_1)^k] = 1 \qquad (11)$$

Equation (11) can be solved numerically for *k*, assuming a specific density for *Z* and given values of $\alpha_1$ and $\beta_1$. Given a sample of returns X$_*$, we can fit a power function $cX_*^{-2k*}$, to the specific tail (e.g. X$_*$ = X > Quantile(X, 0.95)) and compare estimates *k* and $k_*$ of the tail index. In the following Table 5, we present robust estimates of *k* assuming the GARCH(1,1) Student-t model with estimated parameters in Table 1B in the Appendix and estimates of $k_*$ in the right (95%) and left (5%) tail using the empirical distribution of returns.

[INSERT TABLE 5]

The average deviation between the GARCH estimate of the tail index *k* and the empirical estimate *k\** is 3.83% with the quantile 5% and 7.10% with the quantile 95%. Therefore, in most cases, the GARCH model's implied tails are close to the empirical ones, especially with the 5% quantile. This is the reason GARCH-t models capture the tail risk so well. On average *k* is 1.47, higher than *k\** (95%) which is 1.18 and *k* (5%), 1.28. This shows that empirical tails decay more slowly than predicted by GARCH models, especially beyond the 95% quantile. This suggests that the right-hand tail is more difficult to model than the left-hand tail. As a result, GARCH models are more adequate to investors taking long positions in electricity futures, a situation that also applies to other European energy futures markets, as documented in Westgaard et al. (2019a).

6. **Conclusion**

Agents in the deregulated electricity markets are well aware of the high volatility of prices and so have an incentive in implementing risk management policies, for instance, using futures contracts of different maturities. Thus, research on the tail risk associated with the prices of those futures contracts may help market participants in the design and implementation of their hedging and trading strategies. This paper examines some parametric and non-parametric approaches to forecasting VaR and ES in four markets (Norpool, Germany, France, and Spain) in the period 2008-2017 for the futures price of two contracts extensively used by market participants, monthly and yearly contracts.

The empirical evidence suggests that AR(1)-GARCH(1,1) models with t distributions present better performance than historical simulation, EWQR, and quantile regression models when predicting the VaR and ES in all markets and contracts. The reasons explaining this result are the ability of GARCH models of simultaneously capturing volatility clustering and fitting the heavy tails of the distribution of returns. However, the right-hand tail risk is more difficult to model than the left-hand tail. As a result, GARCH models are more adequate to investors taking long positions in electricity futures, a situation that also applies to other European energy futures markets. In addition to that, the performance of GARCH models in the NordPool and German markets is better than in the Spanish and French markets. Therefore, agents taking long positions in the NordPool, and German markets are advised to use GARCH-type models as a way of gauging the tail risk as well as a way of computing regulatory capital requirements and setting the size of default funds. In the Spanish market and especially in the French market, more study is needed before a recommendation can be done.

.

Popular methods used in the industry such as the Historical Simulation method and more innovative methods such as those based on Quantile Regression misestimate the tail risk of electricity futures returns. The reason is that HS models produce a constant VaR forecast, that changes only when very large returns appear. After that, the VaR forecast remains at the new level for the number of days contained in the rolling window. Therefore, the HS method tends, at first, to underestimate risk and, after one big return, to overestimate risk. Furthermore, the VaR forecast generated by the HS model is largely insensitive to volatility clustering. Similar behavior to HS is produced by the EWQR, although this model presents more adaptability to volatility clustering. Finally, the Quantile Regression (QR-COM) model generates almost constant VaR forecasts because the regressors have very low explanatory power, and therefore their changes hardly impact the VaR forecast. The regressors (fuel prices and backward-looking volatility measures) have been applied in other commodity markets with some success, but in this case, their explanatory power turns out to be low. Searching for other explanatory variables useful in the case of electricity futures is an interesting avenue for future research.

The policy recommendation to financial market regulators is that risk weights, and the associated capital ratios to be required to investments in long positions these contracts should be computed using GARCH-type methods. The tail risk in short positions is more difficult to assess, and therefore more conservative capital requirements should be required. Widely used methods such as HS produce lower capital buffers, increasing the risk of financial distress for the incumbent firms. As an added policy recommendation to the administrators of futures markets, in charge of setting margin account regulations, is that, although the volatility of the yearly contract's returns is lower than the volatility of monthly contract returns, forecasting the VaR of the yearly contract is more difficult. Thus, margin requirements should be increased to account for this.

Looking forward, to improve this study, more general GARCH-type models could be studied, allowing more flexible distribution for the innovations such as the skewed-t, Extreme Value Theory-based, and the Normal Inverse Gaussian. We also acknowledge that our empirical results apply for only a one-day-ahead time horizon. Thus, assessing the VaR and ES performance of the models for other time horizons (weekly, monthly) may provide useful information to market participants and regulators. Considering additional backtesting procedures, especially for ES in the line of Du and Escanciano (2017) is another promising area of study, that we left for future research.

# References

Acerbi, C., and D. Tasche, 2002. On the coherence of expected shortfall. *Journal of Banking and Finance*, 26, 1487-1503.

Acerbi, C., and Székely, B., 2014. Backtesting expected shortfall. *Asia Risk*, (Dec 2014/Jan 2015): 42-47.

Andresen, A., Koekebakker, S., and S. Westgaard, S., 2010. Modeling electricity forward prices using the multivariate normal inverse Gaussian distribution. *The Journal of Energy Markets*, 3, 3–25.

Benth, F.E., J. Šaltyte-Benth, and S. Koekebakker (2008). *Stochastic modelling of electricity and related markets*. World Scientific Publishing.

Blanco, I., Peña, J.I, and Rodriguez, R., 2018. Modeling Electricity Swaps with Stochastic Forward Premium Models. *The Energy Journal*, 39,2, 1-33.

Boudoukh, J., M. Richardson, and R. F. Whitelaw., 1998. The Best of Both Worlds. *Risk,* 11, May, 64–67.

Boroumand, R. H., Goutte, S., Porcher, S., and Porcher. T., 2015. Hedging strategies in energy markets: The case of electricity retailers. *Energy Economics*, 51, 503-509.

Bunn, D., Andresen, A., Chen, D., and Westgaard, S., 2016. Analysis and Forecasting of Electricity Price Risk with Quantile Factor Models. *The Energy Journal*, 37,1, 101-122.

Byström, H.N., 2005. Extreme value theory and extremely large electricity price changes. *International Review of Economics and Finance*, 14, 1, 41–55.

Chan, K.F., and Gray, P., 2006. Using extreme value theory to measure value-at-risk for daily electricity spot prices. *International Journal of Forecasting*, 22, 283–300.

Christoffersen, P., 1998. Evaluating Interval Forecasts. *International Economic Review*, 39, 841–862

Commission de Régulation de L'Énergie, 2017. Surveillance Rapport 2016-2017. Le fonctionnement des marchés de gros de l'électricité, du CO2 et du gaz natural. https://www.cre.fr/Documents/Publications/Rapports-thematiques/Rapport-sur-le-fonctionnement-des-marches-de-gros-2016-2017.

CNMC, 2014. Informes de supervisión de mercados a plazo de Energía Eléctrica en España (Diciembre 2013, Enero 2014). Comisión Nacional de los Mercados y la Competencia. https://www.cnmc.es/ambitos-de-actuacion/energia/mercado-electrico.

Dahlgren, R., Liu, C.-C., Lawarree, J., 2003. Risk assessment in energy trading. *IEEE Transactions on Power Systems* 18, 503–511.

Davis, R. A., and Mikosch, T., 2009. Extreme value theory for GARCH processes. *Handbook of Financial Time Series*, 187-200.

Deng, S., Oren, S. 2006. Electricity derivatives and risk management. *Energy*, 31, 940–953

Du, Z., and Escanciano, J. C., 2017. Backtesting Expected Shortfall: Accounting for Tail Risk. *Management Science*, 63, 4, 940-958.

Emmer, S., Kratz, M., and Tasche, D., 2015. What Is the Best Risk Measure in Practice? A Comparison of Standard Measures. *Journal of Risk* 18, 2, 31-60.

Escribano, Á., Peña, J. I., and Villaplana, P., 2011. Modeling Electricity Prices: International Evidence. *Oxford Bulletin of Economics and Statistics*, 73,5, 622- 650.

Eydeland, A., and Wolyniec, K., 2003. *Energy and Power Risk Management*. Wiley Finance.

Fanone, E., Gamba, A., Prokopczuk, M., 2013. The case of negative day-ahead electricity prices. *Energy Economics*, 35, 22-34.

Ferro, C. A. T., and Segers, J., 2003. Inference for clusters of extreme values. *Journal of the Royal Statistical Society, Series B*, 65, 545–556.

Financial Times, 2011a. Germany orders safety review at nuclear plants. https://www.ft.com/content/842c725e-4d81-11e0-85e4-00144feab49a.

Financial Times, 2011b. Merkel attacked over nuclear stance. https://www.ft.com/content/18adf5a8-50a5-11e0-9e89-00144feab49a.

Financial Times, 2018. Trader blows a €100m hole in Nasdaq's Nordic power market. https://www.ft.com/content/43c74e02-b749-11e8-bbc3-ccd7de085ffe.

Hull, J.C., and White, A., 2014. The shortfalls of expected shortfall. *Risk*, November, 10-12.

González-Pedraz, C., Moreno, M., and Peña, J. I., 2014. Tail Risk in Energy portfolios. *Energy Economics*, 46, 422-434.

Haas, M., 2011. New Methods in Backtesting. Financial Engineering, Research Center Caesar, Bonn.

Hagfors, L. I., Bunn D., Kristoffersen, E., Staver, T. T. and Westgaard, S., 2016a. Modeling the UK electricity price distributions using quantile regression. *Energy*, 102, 231-243.

Hagfors, L.I., Kamperud, H. H., Paraschiv, F., Prokopczuk, M., Sator, A., and Westgaard, S., 2016b. Prediction of extreme price occurrences in the German day-ahead electricity market. *Quantitative Finance*, 16, 1929-1948.

Haugom, E., Ray, R., Ullrich, C. J., Veka, S., and Westgaard, S., 2016. A parsimonious quantile regression model to forecast day-ahead value-at-risk. *Finance Research Letters*, 16, 196–207.

Henney, A., Keers, G., 1999. Managing total corporate electricity / energy market risks. *The Electricity Journal*,11, 36–46.

Ioannidis J. P. A., 2005. Why most published research findings are false. *PLoS Med* 2(8): e124.

Jorion, P., 2011. *Financial Risk Manager Handbook*. 6th Edition. Wiley Finance.

Keles, D., Hadzi-Mishev, R., and F. Paraschiv, 2016. Extreme value theory for heavy tails in electricity prices. *Journal of Energy Markets*, 9, 2, 21-50.

Knittel, C. R., and Roberts, M., 2005. An empirical examination of restructured electricity prices. *Energy Economics*, 27, 791–817.

Koenker R., and Bassett G., 1978. Regression quantiles. *Econometrica*, 46, 1, 33-50.

Kupiec, P., 1995. Techniques for Verifying the Accuracy of Risk Management Models. *Journal of Derivatives*, 3, 73–84.

Leong, K., and Siddiqi, R., 1999. Value at risk for power markets. In *Energy Risk Management*, P.C. Fusaro, ed. McGraw Hill.

Liu M., Wu F. F., Ni Y.X., 2006. A survey on risk management in electricity markets. The proceedings of the 2006 IEEE PES General Meeting. Montreal.

Liu, M., and Wu, F. F., 2007. Risk management in a competitive electricity market. *Electrical Power and Energy Systems,* 29, 690-697.

McNeil, A. J., and Frey, R., 2000. Estimation of tail-related risk measures for heteroscedastic financial time series: an extreme value approach. *Journal of Empirical Finance*, 7, 271-300.

McNeil, A. J., Frey, R., and Embrechts, P., 2010. *Quantitative Risk Management: Concepts, Techniques, and Tools*. Princeton University Press.

Mikosch. T. and Starica, C., 2000. Limit theory for the sample autocorrelations and extremes of a GARCH(1, 1) process. *Annals of Statistics*, 28, 1427-1451.

Paraschiv, F., Erni, D., and Pietsch, R., 2014. The impact of renewable energies on EEX day-ahead electricity prices. *Energy Policy*, 73, 196-210.

Paraschiv, F., Hadzi-Mishev, R., and Keles, D., 2016. Extreme Value Theory for Heavy Tails in Electricity Prices. *Journal of Energy Markets*, 9, 21-50.

Pérignon, C., and Smith, D. R., 2010. The level and quality of Value-at-Risk disclosure by commercial banks. *Journal of Banking and Finance*, 34, 362-377.

Pritsker, M., 2006. The hidden dangers of historical simulation. *Journal of Banking and Finance* 30, 561–582.

Quinn, J.A., J.D. Reitzes, J.D., and Schumacher, A. C., 2005. Forward and spot prices in electricity and gas markets. In *Obtaining the Best from Regulation and Competition*, Crew, M.A. and M. Spiegel (Eds.): 109–134. Springer-Verlag.

Sanda, G.E., Tandberg Olsen, E. and Fleten, S.-E., 2013. Selective hedging in hydro-based electricity companies. *Energy Economics*, 40, 326-338.

Taylor, J. W., 2008. Using Exponentially Weighted Quantile Regression to Estimate Value at Risk and Expected Shortfall. *Journal of Financial Econometrics*, 6, 3, 382-406.

Vehviläinen I. and Keppo J. 2003. Managing electricity market price risk. *European Journal of Operational Research* 145, (1), 136–147.

Westgaard, S., Århus, G. H., Frydenberg, M., and Frydenberg, S., 2019a. Value-at-risk in the European energy market: a comparison of parametric, historical simulation, and quantile regression value-at-risk. *The Journal of Risk Model Validation*, 13, 4, 43-69.

Westgaard, S., Paraschiv, F., Ekern, L. L., Naustdal, I., 2019b. Forecasting Price Distributions in the German Electricity Market. *International Financial Markets*, 11-35. Routledge.

Yamai, Y. and Yoshiba T., 2002. Comparative Analysis of Expected Shortfall and Value at Risk: Their Estimation Error, Decomposition, and Optimization. *Monetary and Economic Studies*, January, 87–121

Zhang, S., Chen, H-S, and Pfeiffer, R. M., 2013. A combined p-value test for multiple hypothesis testing. *Journal of Statistical Planning and Inference*, 143, 764-770.

# Tables

## Table 1A: Models

| Model | Short Code | Parameters and Equation |
|---|---|---|
| Historical Simulation | HS | Window size of 250 data points |
| Exponentially Weighted Quantile Regression | EWQR | $\lambda$ = 0.97 and 0.99 |
| Autoregressive model of order 1 and Generalized Autoregressive Conditional Heteroskedastic model of order (1,1) with Student-t innovations with ν degrees of freedom | AR(1)-GARCH(1,1) | $r_j = \phi_0 + \phi_1 r_{j-1} + \sigma_j \varepsilon_j = \mu_j + \sigma_j \varepsilon_j; \quad \varepsilon_j \sim t(\nu)$<br>$\sigma_j^2 = \omega_0 + \omega_1 \sigma_{j-1}^2 + \omega_2 \varepsilon_{j-1}^2$ |
| Autoregressive model of order 1 and Exponential Generalized Autoregressive Conditional Heteroskedastic model of order (1,1) with Student-t innovations with ν degrees of freedom | AR(1)-EGARCH(1,1) | $r_j = \phi_0 + \phi_1 r_{j-1} + \sigma_j \varepsilon_j = \mu_j + \sigma_j \varepsilon_j; \quad \varepsilon_j \sim t(\nu)$<br>$\ln(\sigma_j^2) = \beta_0 + \beta_1 \varepsilon_{j-1} + \beta_2 \ln(\sigma_{j-1}^2) + \beta_3(\varepsilon_{j-1} - E\|\varepsilon_{j-1}\|)$ |
| Autoregressive model of order 1 and Glosten-Jagannathan-Runkle GARCH model of order (1,1) with Student-t innovations with ν degrees of freedom | AR(1)-GJR(1,1) | $r_j = \phi_0 + \phi_1 r_{j-1} + \sigma_j \varepsilon_j = \mu_j + \sigma_j \varepsilon_j; \quad \varepsilon_j \sim t(\nu)$<br>$\sigma_j^2 = \kappa_0 + \kappa_1 \varepsilon_{j-1}^2 + \kappa_2 I_{j-1}(\varepsilon_{j-1} < 0)\varepsilon_{j-1}^2 + \kappa_3 \sigma_{j-1}^2$ |
| Quantile Regression model for quantile α with fuel prices and daily, weekly, and monthly backward-looking volatility measures as explanatory variables | QR-COM | $Q_\alpha(r_j) = \omega_{\alpha,0} + \sum_{m=1}^{M} \delta_{\alpha,m} VolX_{j,m} + \sum_{k=1}^{K} \beta_{\alpha,k} FuelPX_{j,k} + \eta_{\alpha,j}$ |

## Table 1B: VaR and ES Backtesting methods

| Method | Risk Measure | Short Code | Comments |
|---|---|---|---|
| Binomial test (Jorion, 2011) | VaR | Bin | The Bin assesses if the number of failures is consistent with the VaR confidence level. |
| The proportion of failures test (Kupiec, 1995) | VaR | POF | The POF tests if the proportion of failures is consistent with the VaR confidence level |
| Conditional coverage independence test (Christoffersen, 1998) | VaR | CCI | The CCI assesses the independence of failures on consecutive time periods. |
| Dynamic Quantile test (Engle and Manganelli, 2004) | VaR | DQ | The DQ evaluates the existence of a correlation between failures, their lagged values, and the contemporaneous VaR estimate |
| Unconditional Test under the Normal distribution (Acerbi and Székely, 2014) | ES | UN-N | The UN-Normal tests if the average of the ES-scaled losses beyond VaR is consistent with the tails of a Normal distribution |
| Unconditional Test under the Student-t distribution (Acerbi and Székely, 2014) | ES | UN-t | The UN-t tests if the average of the ES-scaled losses beyond VaR is consistent with the tails of a Student-t distribution with 3 degrees of freedom |

## Table 1C: Futures contracts code and sample period

| Country | Short Code | Market | Long Code | Sample |
|---|---|---|---|---|
| Germany | F1BM | EEX | Phelix-DE/AT Base Month Future | 02.01.2008-25.09.2017 |
| Germany | F1BY | EEX | Phelix-DE/AT Base Year Future | 02.01.2008-25.09.2017 |
| France | F2BM | EEX | French Base Month Future | 02.01.2008-23.11.2011 |
| France | F2BY | EEX | French Basel Year Future | 02.01.2008-23.11.2011 |
| France | F7BM | EEX | French Base Month Future | 24.11.2011-25.09.2017 |
| France | F7BY | EEX | French Base Year Future | 24.11.2011-25.09.2017 |
| Spain | FEBM | EEX | Spanish Base Month Future | 01.04.2014-25.09.2017 |
| Spain | FEBY | EEX | Spanish Base Year Future | 01.04.2014-25.09. 2017 |
| Spain | FTBM | OMIP | Spanish Base Month Future | 02.01.2008-31.03.2014 |
| Spain | FTBY | OMIP | Spanish Base Year Future | 02.01.2008-31.03.2014 |
| NordPool | ENOYC1 | Nasdaq OMX | Oslo ASA Euro ENO Yearly Contract Energy Future | 02.01.2008-25.09.2017 |
| NordPool | ENOMC1 | Nasdaq OMX | Oslo ASA Euro ENO Monthly Contract Energy Future | 02.01.2008-25.09.2017 |

## Table 2: Descriptive statistics returns

This table reports descriptive statistics of the eight series of futures price returns. The sample period is 02.01.2008-25.09.2017. JB is the p-value of the Jarque-Bera normality test. Q is the Ljung-Box test with 10 and 20 lags and pvQ is the corresponding p-value

| Panel A: Full Sample | | | | | | | | |
|---|---|---|---|---|---|---|---|---|
| | M1 | | | | Y1 | | | |
| | FR | GE | SP | NP | FR | GE | SP | NP |
| Mean (%) | -0.060% | -0.074% | -0.137% | -0.082% | -0.014% | -0.027% | -0.020% | -0.016% |
| Std_dev | 0.026 | 0.016 | 0.017 | 0.029 | 0.011 | 0.011 | 0.007 | 0.015 |
| Minimum | -0.263 | -0.122 | -0.134 | -0.167 | -0.132 | -0.059 | -0.050 | -0.090 |
| Maximum | 0.202 | 0.163 | 0.162 | 0.215 | 0.089 | 0.065 | 0.055 | 0.099 |
| Skewness | -0.333 | 0.275 | -0.020 | 0.193 | -0.481 | 0.179 | -0.271 | -0.006 |
| Kurtosis | 15.789 | 10.670 | 18.539 | 7.329 | 16.873 | 7.203 | 10.978 | 7.187 |
| Quantile 1 | -0.063 | -0.045 | -0.050 | -0.079 | -0.033 | -0.030 | -0.022 | -0.044 |
| Quantile 5 | -0.038 | -0.026 | -0.026 | -0.046 | -0.016 | -0.017 | -0.012 | -0.025 |
| Quantile 95 | 0.035 | 0.024 | 0.021 | 0.043 | 0.017 | 0.016 | 0.011 | 0.023 |
| Quantile 99 | 0.082 | 0.044 | 0.042 | 0.080 | 0.030 | 0.030 | 0.022 | 0.044 |
| JB_Test | 0.001 | 0.001 | 0.001 | 0.001 | 0.001 | 0.001 | 0.001 | 0.001 |
| Q_10 | 149.240 | 85.987 | 59.872 | 37.454 | 50.441 | 33.925 | 65.513 | 11.732 |
| pvQ_10 | 0.000 | 0.000 | 0.000 | 0.000 | 0.000 | 0.000 | 0.000 | 0.303 |
| Q_20 | 176.440 | 96.031 | 90.332 | 47.375 | 98.801 | 61.210 | 79.199 | 24.946 |
| pvQ_20 | 0.000 | 0.000 | 0.000 | 0.001 | 0.000 | 0.000 | 0.000 | 0.204 |
| obsv | 2469 | 2470 | 2463 | 2438 | 2469 | 2470 | 2463 | 2444 |

| Panel B: In-sample | | | | | | | | |
|---|---|---|---|---|---|---|---|---|
| | M1 | | | | Y1 | | | |
| | FR | GE | SP | NP | FR | GE | SP | NP |
| Mean (%) | -0.099% | -0.135% | -0.177% | -0.127% | -0.027% | -0.048% | -0.032% | -0.026% |
| Std_dev | 0.023 | 0.016 | 0.017 | 0.028 | 0.010 | 0.010 | 0.008 | 0.015 |
| Minimum | -0.131 | -0.122 | -0.133 | -0.167 | -0.056 | -0.059 | -0.050 | -0.090 |
| Maximum | 0.202 | 0.163 | 0.162 | 0.127 | 0.060 | 0.065 | 0.055 | 0.092 |
| Skewness | 0.441 | 0.346 | 0.011 | -0.047 | -0.006 | 0.184 | -0.228 | -0.109 |
| Kurtosis | 10.006 | 13.132 | 19.356 | 5.759 | 7.912 | 8.677 | 10.188 | 7.318 |
| Quantile 1 | -0.061 | -0.044 | -0.052 | -0.080 | -0.033 | -0.030 | -0.026 | -0.044 |
| Quantile 5 | -0.038 | -0.026 | -0.026 | -0.046 | -0.015 | -0.015 | -0.014 | -0.024 |
| Quantile 95 | 0.034 | 0.023 | 0.020 | 0.043 | 0.015 | 0.015 | 0.012 | 0.023 |
| Quantile 99 | 0.068 | 0.043 | 0.041 | 0.076 | 0.029 | 0.029 | 0.025 | 0.044 |
| JB_Test | 0.001 | 0.001 | 0.001 | 0.001 | 0.001 | 0.001 | 0.001 | 0.001 |
| Q_10 | 48.488 | 57.228 | 44.393 | 27.964 | 34.202 | 37.535 | 44.985 | 18.066 |
| pvQ_10 | 0.000 | 0.000 | 0.000 | 0.002 | 0.000 | 0.000 | 0.000 | 0.054 |
| Q_20 | 52.020 | 65.288 | 88.838 | 38.979 | 49.374 | 51.663 | 56.098 | 29.137 |
| pvQ_20 | 0.000 | 0.000 | 0.000 | 0.007 | 0.000 | 0.000 | 0.000 | 0.085 |
| obsv | 1771 | 1772 | 1762 | 1756 | 1771 | 1772 | 1762 | 1756 |

| Panel C: Out-of-sample | | | | | | | | |
|---|---|---|---|---|---|---|---|---|
| | M1 | | | | Y1 | | | |
| | FR | GE | SP | NP | FR | GE | SP | NP |
| Mean (%) | 0.039% | 0.078% | -0.036% | 0.032% | 0.019% | 0.026% | 0.011% | 0.008% |
| Std_dev | 0.031 | 0.017 | 0.016 | 0.030 | 0.014 | 0.012 | 0.005 | 0.016 |
| Minimum | -0.263 | -0.074 | -0.134 | -0.146 | -0.132 | -0.050 | -0.030 | -0.063 |
| Maximum | 0.167 | 0.081 | 0.124 | 0.215 | 0.089 | 0.049 | 0.027 | 0.099 |
| Skewness | -1.147 | 0.097 | -0.093 | 0.695 | -0.995 | 0.122 | -0.328 | 0.238 |
| Kurtosis | 18.005 | 5.641 | 16.510 | 10.334 | 21.909 | 5.000 | 7.716 | 6.854 |
| Quantile 1 | -0.089 | -0.050 | -0.040 | -0.073 | -0.034 | -0.031 | -0.015 | -0.044 |
| Quantile 5 | -0.041 | -0.025 | -0.026 | -0.047 | -0.019 | -0.020 | -0.008 | -0.025 |
| Quantile 95 | 0.042 | 0.028 | 0.021 | 0.043 | 0.020 | 0.020 | 0.009 | 0.023 |
| Quantile 99 | 0.095 | 0.049 | 0.049 | 0.087 | 0.036 | 0.034 | 0.015 | 0.044 |
| JB_Test | 0.001 | 0.001 | 0.001 | 0.001 | 0.001 | 0.001 | 0.001 | 0.001 |
| Q_10 | 112.000 | 38.868 | 31.032 | 21.230 | 50.212 | 17.240 | 32.941 | 12.313 |
| pvQ_10 | 0.000 | 0.000 | 0.001 | 0.020 | 0.000 | 0.069 | 0.000 | 0.265 |
| Q_20 | 147.920 | 55.229 | 39.324 | 40.272 | 102.470 | 38.016 | 47.028 | 30.761 |
| pvQ_20 | 0.000 | 0.000 | 0.006 | 0.005 | 0.000 | 0.009 | 0.001 | 0.058 |
| obsv | 698 | 698 | 701 | 688 | 698 | 698 | 701 | 688 |

# Table 3: Summary of backtesting results

This table reports a summary of the backtesting results. The table shows the number of times and percentages that data do not reject the overall Fisher test at 1% level for all VaR models, contracts, and countries. GT is the total for both tails and %GT is the proportion of no rejections (e.g. 0.13 means that in 13 out of 100 cases the Fisher test does not reject the null of model adequacy). GTL is the total for the left tail (5%, 1%). GTR is the total for the right tail (95%, 99%). With ES, we present the number of times and proportions UN-N and UN-t test does not reject the null of model adequacy. We check the consistency of the tests' outcomes.

**Panel A1: VaR (All contracts)**

|  | Grand Total | % GT | GTL | %GTL | GTR | %GTR |
|---|---|---|---|---|---|---|
| HS | 7 | 0.22 | 5 | 0.31 | 2 | 0.13 |
| AR(1)-GARCH(1,1) | 28 | 0.88 | 15 | 0.94 | 13 | 0.81 |
| AR(1)-EGARCH(1,1) | 26 | 0.81 | 13 | 0.81 | 13 | 0.81 |
| AR(1)-GJR(1,1) | 24 | 0.75 | 13 | 0.81 | 11 | 0.69 |
| EWQR | 14 | 0.44 | 9 | 0.56 | 5 | 0.31 |
| QR-COM | 9 | 0.28 | 9 | 0.56 | 0 | 0.00 |

**Panel B1: ES (All contracts)**

|  | Grand Total | % GT | GTL | %GTL | GTR | %GTR |
|---|---|---|---|---|---|---|
| HS | 40 | 0.63 | 25 | 0.78 | 15 | 0.47 |
| AR(1)-GARCH(1,1) | 64 | 1.00 | 32 | 1.00 | 32 | 1.00 |
| AR(1)-EGARCH(1,1) | 64 | 1.00 | 32 | 1.00 | 32 | 1.00 |
| AR(1)-GJR(1,1) | 64 | 1.00 | 32 | 1.00 | 32 | 1.00 |
| EWQR | 40 | 0.63 | 24 | 0.75 | 16 | 0.50 |
| QR-COM | 21 | 0.33 | 18 | 0.56 | 3 | 0.09 |

**Panel A2: VaR (M1)**

|  | Grand Total | %GT | GTL | %GTL | GTR | %GTR |
|---|---|---|---|---|---|---|
| HS | 4 | 0.25 | 4 | 0.50 | 0 | 0.00 |
| AR(1)-GARCH(1,1) | 14 | 0.88 | 8 | 1.00 | 6 | 0.75 |
| AR(1)-EGARCH(1,1) | 13 | 0.81 | 7 | 0.88 | 6 | 0.75 |
| AR(1)-GJR(1,1) | 13 | 0.81 | 8 | 1.00 | 5 | 0.63 |
| EWQR | 9 | 0.56 | 6 | 0.75 | 3 | 0.38 |
| QR-COM | 6 | 0.38 | 6 | 0.75 | 0 | 0.00 |

**Panel B2: VaR (Y1)**

|  | Grand Total | %GT | GTL | %GTL | GTR | %GTR |
|---|---|---|---|---|---|---|
| HS | 3 | 0.19 | 1 | 0.13 | 2 | 0.25 |
| AR(1)-GARCH(1,1) | 14 | 0.88 | 7 | 0.88 | 7 | 0.88 |
| AR(1)-EGARCH(1,1) | 13 | 0.81 | 6 | 0.75 | 7 | 0.88 |
| AR(1)-GJR(1,1) | 11 | 0.69 | 5 | 0.63 | 6 | 0.75 |
| EWQR | 5 | 0.31 | 3 | 0.38 | 2 | 0.25 |
| QR-COM | 3 | 0.19 | 3 | 0.38 | 0 | 0.00 |

**Panel C2: ES (M1)**

|  | Grand Total | %GT | GTL | %GTL | GTR | %GTR |
|---|---|---|---|---|---|---|
| HS | 28 | 0.88 | 15 | 0.94 | 13 | 0.81 |
| AR(1)-GARCH(1,1) | 32 | 1.00 | 16 | 1.00 | 16 | 1.00 |
| AR(1)-EGARCH(1,1) | 32 | 1.00 | 16 | 1.00 | 16 | 1.00 |
| AR(1)-GJR(1,1) | 32 | 1.00 | 16 | 1.00 | 16 | 1.00 |
| EWQR | 32 | 1.00 | 16 | 1.00 | 16 | 1.00 |
| QR-COM | 14 | 0.44 | 12 | 0.75 | 2 | 0.13 |

**Panel D2: ES(Y1)**

|  | Grand Total | %GT | GTL | %GTL | GTR | %GTR |
|---|---|---|---|---|---|---|
| HS | 12 | 0.38 | 10 | 0.63 | 2 | 0.13 |
| AR(1)-GARCH(1,1) | 32 | 1.00 | 16 | 1.00 | 16 | 1.00 |
| AR(1)-EGARCH(1,1) | 32 | 1.00 | 16 | 1.00 | 16 | 1.00 |
| AR(1)-GJR(1,1) | 32 | 1.00 | 16 | 1.00 | 16 | 1.00 |
| EWQR | 8 | 0.25 | 8 | 0.50 | 0 | 0.00 |
| QR-COM | 7 | 0.22 | 6 | 0.38 | 1 | 0.06 |

**Panel E1: VAR (Markets)**

|               | %GT NP | %GT GE | %GT SP | %GT FR |
|---------------|--------|--------|--------|--------|
| HS            | 0.00   | 0.38   | 0.38   | 0.13   |
| AR(1)-GARCH(1,1) | 1.00 | 1.00   | 0.88   | 0.63   |
| AR(1)-EGARCH(1,1) | 1.00 | 1.00  | 0.75   | 0.50   |
| AR(1)-GJR(1,1) | 1.00  | 0.88   | 0.75   | 0.38   |
| EWQR          | 0.50   | 0.50   | 0.50   | 0.25   |
| QR-COM        | 0.38   | 0.38   | 0.38   | 0.00   |

**Panel E2: ES (Markets)**

|               | %GT NP | %GT GE | %GT SP | %GT FR |
|---------------|--------|--------|--------|--------|
| HS            | 0.69   | 0.63   | 0.69   | 0.50   |
| AR(1)-GARCH(1,1) | 1.00 | 1.00   | 1.00   | 1.00   |
| AR(1)-EGARCH(1,1) | 1.00 | 1.00  | 1.00   | 1.00   |
| AR(1)-GJR(1,1) | 1.00  | 1.00   | 1.00   | 1.00   |
| EWQR          | 0.75   | 0.75   | 0.50   | 0.50   |
| QR-COM        | 0.25   | 0.38   | 0.44   | 0.25   |

# Table 4: Out-of-sample VaR violations for each model.

The out-of-sample number of observations is 698 (FR, GE), 701(SP), and 688 (NP). A positive (negative) Deviation implies risk underestimation (overestimation).

| Panel A: Nordpool. M1 and Y1 | | M1 | | | Y1 | | |
|---|---|---|---|---|---|---|---|
| | Quantile | Expected | Failures | Deviation (F-E)/E | Expected | Failures | Deviation (F-E)/E |
| **HS** | | | | **41.50%** | | | **69.33%** |
| | 1% | 6.82 | 14 | 105.28% | 6.88 | 14 | 103.49% |
| | 5% | 34.1 | 38 | 11.44% | 34.4 | 45 | 30.81% |
| | 95% | 34.1 | 35 | 2.64% | 34.4 | 48 | 39.53% |
| | 99% | 6.82 | 10 | 46.63% | 6.88 | 14 | 103.49% |
| **AR(1)-GARCH(1,1)** | | | | **10.70%** | | | **13.37%** |
| | 1% | 6.82 | 9 | 31.96% | 6.88 | 10 | 45.35% |
| | 5% | 34.1 | 32 | -6.16% | 34.4 | 31 | -9.88% |
| | 95% | 34.1 | 34 | -0.29% | 34.4 | 40 | 16.28% |
| | 99% | 6.82 | 8 | 17.30% | 6.88 | 7 | 1.74% |
| **AR(1)-EGARCH(1,1)** | | | | **17.30%** | | | **9.01%** |
| | 1% | 6.82 | 10 | 46.63% | 6.88 | 8 | 16.28% |
| | 5% | 34.1 | 35 | 2.64% | 34.4 | 35 | 1.74% |
| | 95% | 34.1 | 35 | 2.64% | 34.4 | 40 | 16.28% |
| | 99% | 6.82 | 8 | 17.30% | 6.88 | 7 | 1.74% |
| **AR(1)-GJR(1,1)** | | | | **-1.03%** | | | **-12.79%** |
| | 1% | 6.82 | 8 | 17.30% | 6.88 | 6 | -12.79% |
| | 5% | 34.1 | 30 | -12.02% | 34.4 | 28 | -18.60% |
| | 95% | 34.1 | 30 | -12.02% | 34.4 | 37 | 7.56% |
| | 99% | 6.82 | 7 | 2.64% | 6.88 | 5 | -27.33% |
| **EWQR** | | | | **29.03%** | | | **32.99%** |
| | 1% | 6.82 | 11 | 61.29% | 6.88 | 9 | 30.81% |
| | 5% | 34.1 | 33 | -3.23% | 34.4 | 40 | 16.28% |
| | 95% | 34.1 | 38 | 11.44% | 34.4 | 43 | 25.00% |
| | 99% | 6.82 | 10 | 46.63% | 6.88 | 11 | 59.88% |
| **QR-COM** | | | | **80.35%** | | | **127.47%** |
| | 1% | 6.82 | 8 | 17.30% | 6.88 | 32 | 365.12% |
| | 5% | 34.1 | 32 | -6.16% | 34.4 | 39 | 13.37% |
| | 95% | 34.1 | 84 | 146.33% | 34.4 | 64 | 86.05% |
| | 99% | 6.82 | 18 | 163.93% | 6.88 | 10 | 45.35% |

| Panel B: Germany. M1 and Y1 | | M1 | | | Y1 | | |
|---|---|---|---|---|---|---|---|
| | Quantile | Expected | Failures | Deviation (F-E)/E | Expected | Failures | Deviation (F-E)/E |
| **HS** | | | | **41.83%** | | | **76.22%** |
| | 1% | 6.98 | 9 | 28.94% | 6.98 | 13 | 86.25% |
| | 5% | 34.9 | 36 | 3.15% | 34.9 | 53 | 51.86% |
| | 95% | 34.9 | 42 | 20.34% | 34.9 | 53 | 51.86% |
| | 99% | 6.98 | 15 | 114.90% | 6.98 | 15 | 114.90% |
| **AR(1)-GARCH(1,1)** | | | | **-10.46%** | | | **-14.04%** |
| | 1% | 6.98 | 2 | -71.35% | 6.98 | 5 | -28.37% |
| | 5% | 34.9 | 28 | -19.77% | 34.9 | 31 | -11.17% |
| | 95% | 34.9 | 47 | 34.67% | 34.9 | 39 | 11.75% |
| | 99% | 6.98 | 8 | 14.61% | 6.98 | 5 | -28.37% |
| **AR(1)-EGARCH(1,1)** | | | | **-10.46%** | | | **-15.47%** |
| | 1% | 6.98 | 2 | -71.35% | 6.98 | 7 | 0.29% |
| | 5% | 34.9 | 30 | -14.04% | 34.9 | 31 | -11.17% |
| | 95% | 34.9 | 45 | 28.94% | 34.9 | 37 | 6.02% |
| | 99% | 6.98 | 8 | 14.61% | 6.98 | 3 | -57.02% |
| **AR(1)-GJR(1,1)** | | | | **-15.47%** | | | **-14.76%** |
| | 1% | 6.98 | 2 | -71.35% | 6.98 | 5 | -28.37% |
| | 5% | 34.9 | 26 | -25.50% | 34.9 | 30 | -14.04% |
| | 95% | 34.9 | 42 | 20.34% | 34.9 | 39 | 11.75% |
| | 99% | 6.98 | 8 | 14.61% | 6.98 | 5 | -28.37% |
| **EWQR** | | | | **28.22%** | | | **54.01%** |
| | 1% | 6.98 | 6 | -14.04% | 6.98 | 11 | 57.59% |
| | 5% | 34.9 | 32 | -8.31% | 34.9 | 44 | 26.07% |
| | 95% | 34.9 | 42 | 20.34% | 34.9 | 46 | 31.81% |
| | 99% | 6.98 | 15 | 114.90% | 6.98 | 14 | 100.57% |
| **QR-COM** | | | | **71.92%** | | | **307.59%** |
| | 1% | 6.98 | 8 | 14.61% | 6.98 | 10 | 43.27% |
| | 5% | 34.9 | 33 | -5.44% | 34.9 | 48 | 37.54% |
| | 95% | 34.9 | 47 | 34.67% | 34.9 | 101 | 189.40% |
| | 99% | 6.98 | 24 | 243.84% | 6.98 | 74 | 960.17% |

| Panel C: Spain M1 and Y1 | | M1 | | | Y1 | | |
|---|---|---|---|---|---|---|---|
| | Quantile | Expected | Failures | Deviation (F-E)/E | Expected | Failures | Deviation (F-E)/E |
| **HS** | | | | **13.41%** | | | **34.09%** |
| | 1% | 7.01 | 9 | 28.39% | 7.01 | 12 | 71.18% |
| | 5% | 35.05 | 36 | 2.71% | 35.05 | 39 | 11.27% |
| | 95% | 35.05 | 33 | -5.85% | 35.05 | 39 | 11.27% |
| | 99% | 7.01 | 9 | 28.39% | 7.01 | 10 | 42.65% |
| **AR(1)-GARCH(1,1)** | | | | **9.84%** | | | **-13.69%** |
| | 1% | 7.01 | 6 | -14.41% | 7.01 | 6 | -14.41% |
| | 5% | 35.05 | 26 | -25.82% | 35.05 | 34 | -3.00% |
| | 95% | 35.05 | 58 | 65.48% | 35.05 | 42 | 19.83% |
| | 99% | 7.01 | 8 | 14.12% | 7.01 | 3 | -57.20% |
| **AR(1)-EGARCH(1,1)** | | | | **2.00%** | | | **-28.67%** |
| | 1% | 7.01 | 5 | -28.67% | 7.01 | 4 | -42.94% |
| | 5% | 35.05 | 24 | -31.53% | 35.05 | 30 | -14.41% |
| | 95% | 35.05 | 54 | 54.07% | 35.05 | 40 | 14.12% |
| | 99% | 7.01 | 8 | 14.12% | 7.01 | 2 | -71.47% |
| **AR(1)-GJR(1,1)** | | | | **5.56%** | | | **-3.00%** |
| | 1% | 7.01 | 6 | -14.41% | 7.01 | 6 | -14.41% |
| | 5% | 35.05 | 24 | -31.53% | 35.05 | 40 | 14.12% |
| | 95% | 35.05 | 54 | 54.07% | 35.05 | 51 | 45.51% |
| | 99% | 7.01 | 8 | 14.12% | 7.01 | 3 | -57.20% |
| **EWQR** | | | | **16.26%** | | | **28.39%** |
| | 1% | 7.01 | 9 | 28.39% | 7.01 | 9 | 28.39% |
| | 5% | 35.05 | 33 | -5.85% | 35.05 | 40 | 14.12% |
| | 95% | 35.05 | 35 | -0.14% | 35.05 | 40 | 14.12% |
| | 99% | 7.01 | 10 | 42.65% | 7.01 | 11 | 56.92% |
| **QR-COM** | | | | **82.60%** | | | **31.24%** |
| | 1% | 7.01 | 6 | -14.41% | 7.01 | 10 | 42.65% |
| | 5% | 35.05 | 25 | -28.67% | 35.05 | 63 | 79.74% |
| | 95% | 35.05 | 51 | 45.51% | 35.05 | 56 | 59.77% |
| | 99% | 7.01 | 30 | 327.96% | 7.01 | 3 | -57.20% |

| Panel D: France. M1 and Y1 | | M1 | | | Y1 | | |
|---|---|---|---|---|---|---|---|
| | Quantile | Expected | Failures | Deviation (F-E)/E | Expected | Failures | Deviation (F-E)/E |
| **HS** | | | | **29.66%** | | | **86.25%** |
| | 1% | 6.98 | 10 | 43.27% | 6.98 | 16 | 129.23% |
| | 5% | 34.9 | 31 | -11.17% | 34.9 | 46 | 31.81% |
| | 95% | 34.9 | 40 | 14.61% | 34.9 | 59 | 69.05% |
| | 99% | 6.98 | 12 | 71.92% | 6.98 | 15 | 114.90% |
| **AR(1)-GARCH(1,1)** | | | | **-7.59%** | | | **28.22%** |
| | 1% | 6.98 | 5 | -28.37% | 6.98 | 9 | 28.94% |
| | 5% | 34.9 | 27 | -22.64% | 34.9 | 36 | 3.15% |
| | 95% | 34.9 | 52 | 49.00% | 34.9 | 38 | 8.88% |
| | 99% | 6.98 | 5 | -28.37% | 6.98 | 12 | 71.92% |
| **AR(1)-EGARCH(1,1)** | | | | **16.05%** | | | **31.09%** |
| | 1% | 6.98 | 6 | -14.04% | 6.98 | 10 | 43.27% |
| | 5% | 34.9 | 31 | -11.17% | 34.9 | 38 | 8.88% |
| | 95% | 34.9 | 51 | 46.13% | 34.9 | 40 | 14.61% |
| | 99% | 6.98 | 10 | 43.27% | 6.98 | 11 | 57.59% |
| **AR(1)-GJR(1,1)** | | | | **46.85%** | | | **-1.15%** |
| | 1% | 6.98 | 6 | -14.04% | 6.98 | 6 | -14.04% |
| | 5% | 34.9 | 43 | 23.21% | 34.9 | 30 | -14.04% |
| | 95% | 34.9 | 57 | 63.32% | 34.9 | 33 | -5.44% |
| | 99% | 6.98 | 15 | 114.90% | 6.98 | 9 | 28.94% |
| **EWQR** | | | | **26.79%** | | | **73.35%** |
| | 1% | 6.98 | 10 | 43.27% | 6.98 | 15 | 114.90% |
| | 5% | 34.9 | 30 | -14.04% | 34.9 | 49 | 40.40% |
| | 95% | 34.9 | 42 | 20.34% | 34.9 | 48 | 37.54% |
| | 99% | 6.98 | 11 | 57.59% | 6.98 | 14 | 100.57% |
| **QR-COM** | | | | **198.71%** | | | **145.70%** |
| | 1% | 6.98 | 15 | 114.90% | 6.98 | 18 | 157.88% |
| | 5% | 34.9 | 122 | 249.57% | 34.9 | 39 | 11.75% |
| | 95% | 34.9 | 55 | 57.59% | 34.9 | 79 | 126.36% |
| | 99% | 6.98 | 33 | 372.78% | 6.98 | 27 | 286.82% |

**Panel E Averages**

| | M1 | Y1 |
|---|---|---|
| **HS** | 34.56% | 62.24% |
| **AR(1)-GARCH(1,1)** | 5.20% | -0.25% |
| **AR(1)-EGARCH(1,1)** | 6.54% | -6.53% |
| **AR(1)-GJR(1,1)** | -2.99% | -10.83% |
| **EWQR** | 25.64% | 37.10% |
| **QR-COM** | 78.81% | 148.44% |

**Table 5: Tail Index GARCH and Empirical**

| Tail Index k | FR | | GE | | SP | | NP | | Average |
|---|---|---|---|---|---|---|---|---|---|
| | M1 | Y1 | M1 | Y1 | M1 | Y1 | M1 | Y1 | |
| $k^*$ (Quantile 95%) | 1.1 | 1.13 | 1.3 | 1.4 | 1.04 | 1.11 | 1.21 | 1.17 | 1.18 |
| $k^*$ (Quantile 5%) | 1.45 | 1.09 | 1.6 | 1.07 | 1.07 | 1.29 | 1.38 | 1.31 | 1.28 |
| GARCH(1,1) Student-t $k$ | 1.61 | 1.01 | 1.55 | 1.13 | 1.01 | 1.03 | 1.76 | 2.62 | 1.47 |
| log($k/k^*$) (5%) | 4.55% | -3.31% | -1.38% | 2.37% | -2.51% | -9.78% | 10.56% | 30.10% | 3.83% |
| log($k/k^*$) (95%) | 16.54% | -4.88% | 7.64% | -9.30% | -1.27% | -3.25% | 16.27% | 35.01% | 7.10% |

**Figure 1**: Prices of electricity futures (€/MWh) of M1 and Y1 contracts for the French (FR), German (GE), Spanish (SP), and NordPool (NP) markets. The sample period goes from January 2, 2008, to September 25, 2017.

Panel A: M1 Contracts

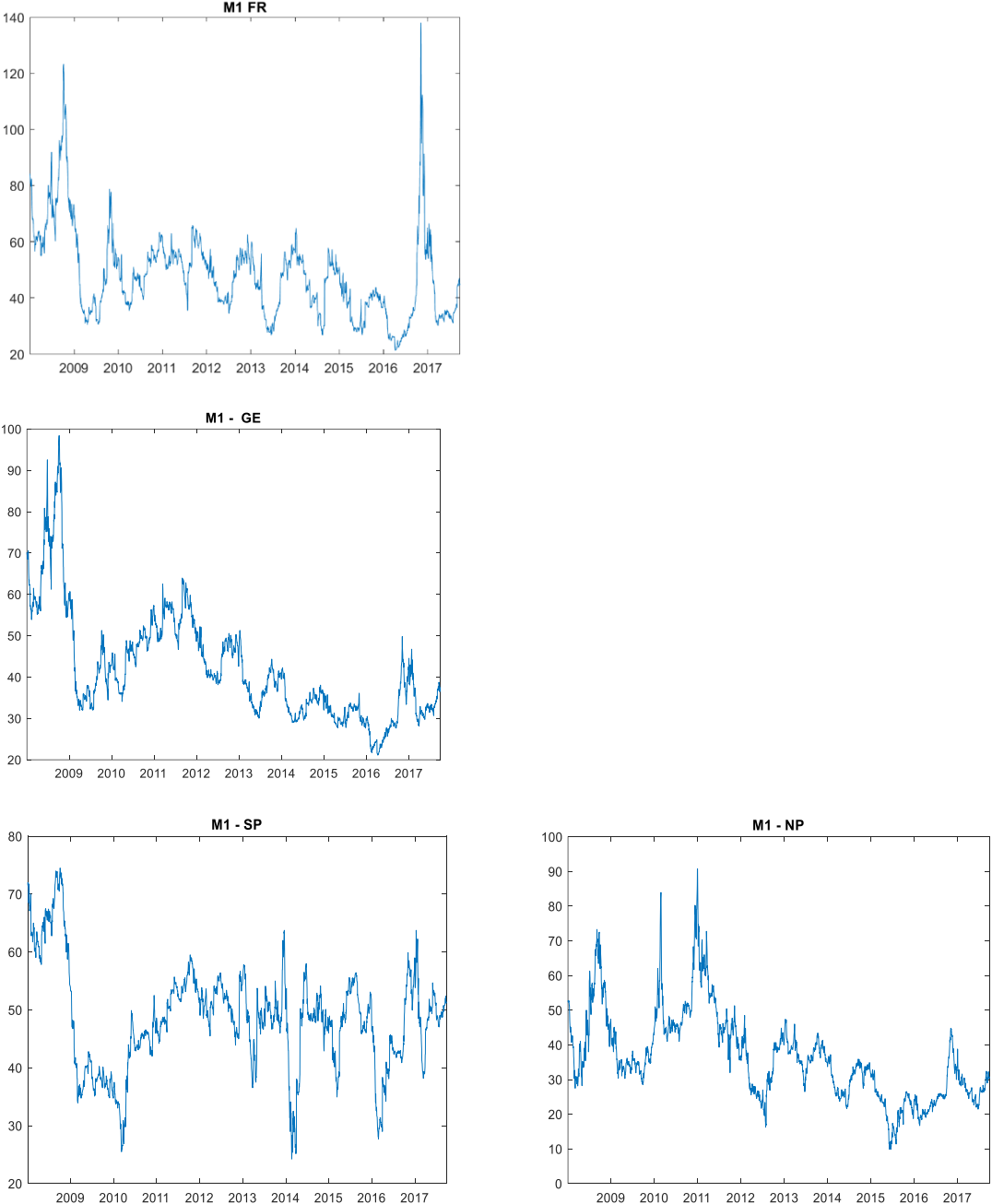

Panel B: Y1 Contracts

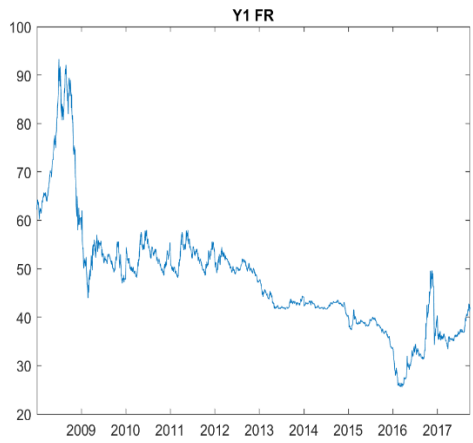
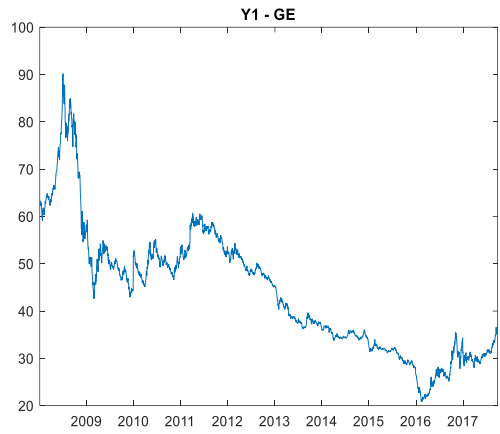
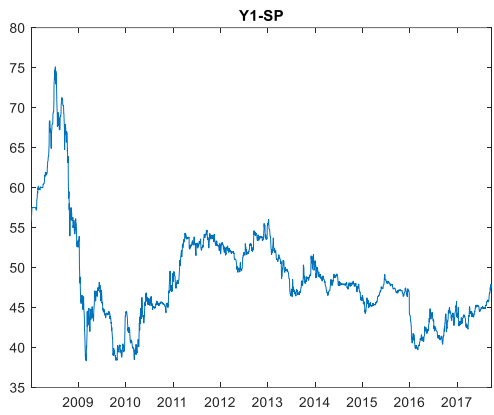
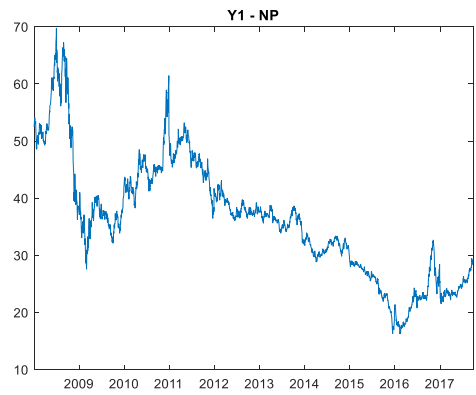

**Figure 2**: Returns of electricity futures of M1 and Y1 contracts for the French (FR), German (GE), Spanish (SP), and NordPool (NP) markets. The sample period goes from January 2, 2008, to September 25, 2017.

Panel A: M1 Contracts

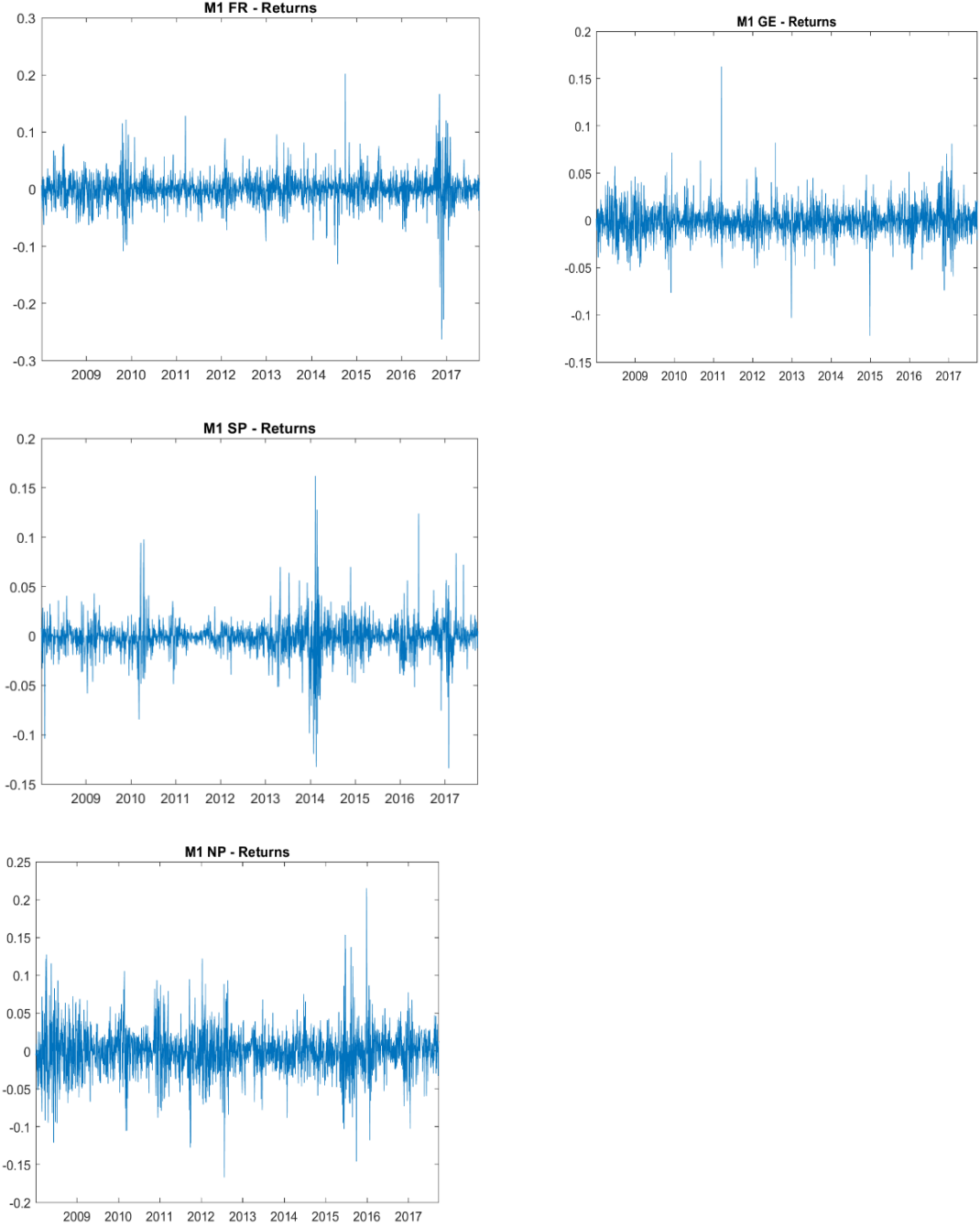

Panel B: Y1 Contracts

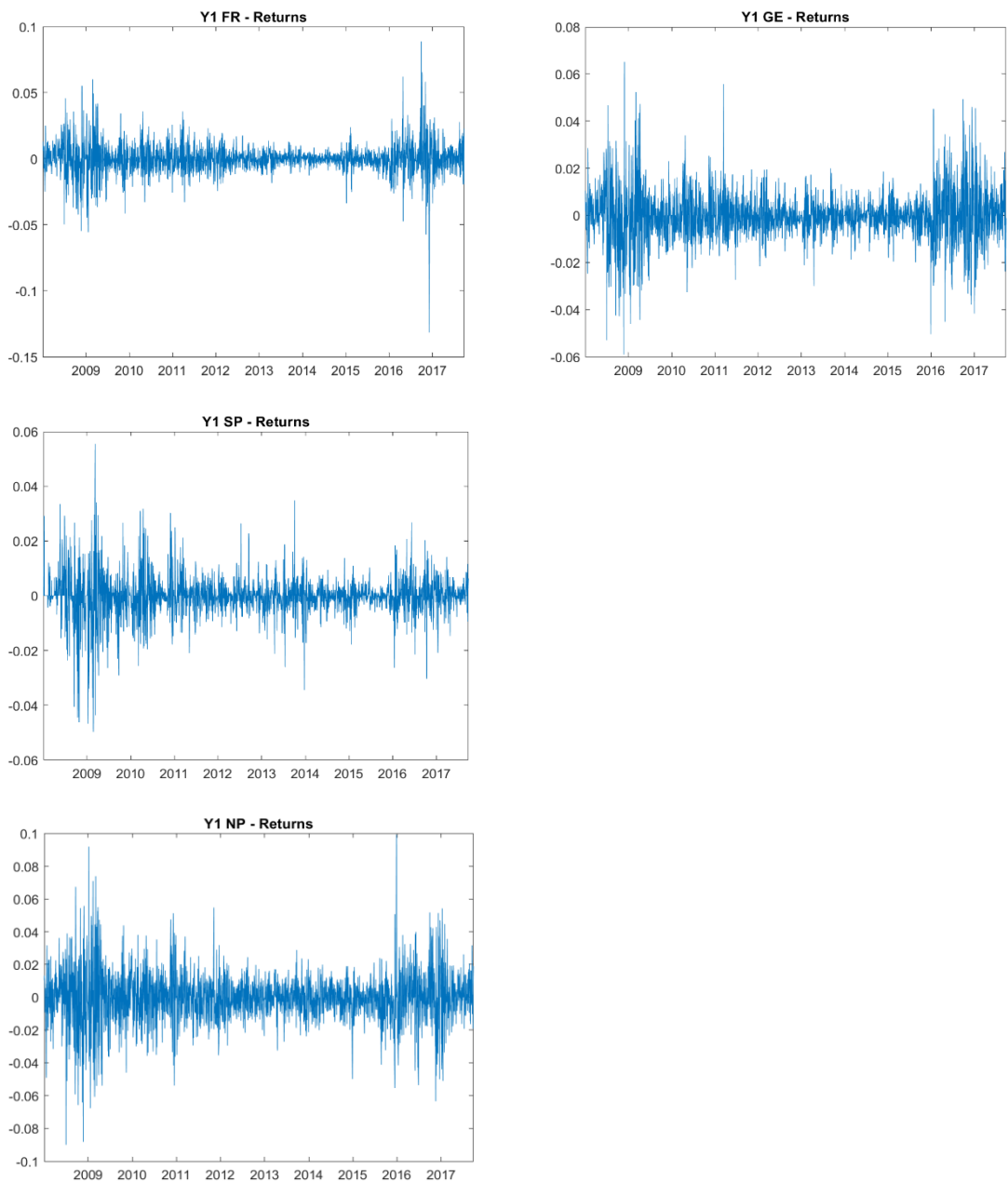

**Figure 3**: Volatility, Skewness, and Kurtosis of daily returns in all markets by year.

Sample period January 2, 2008 – September 25, 2017.

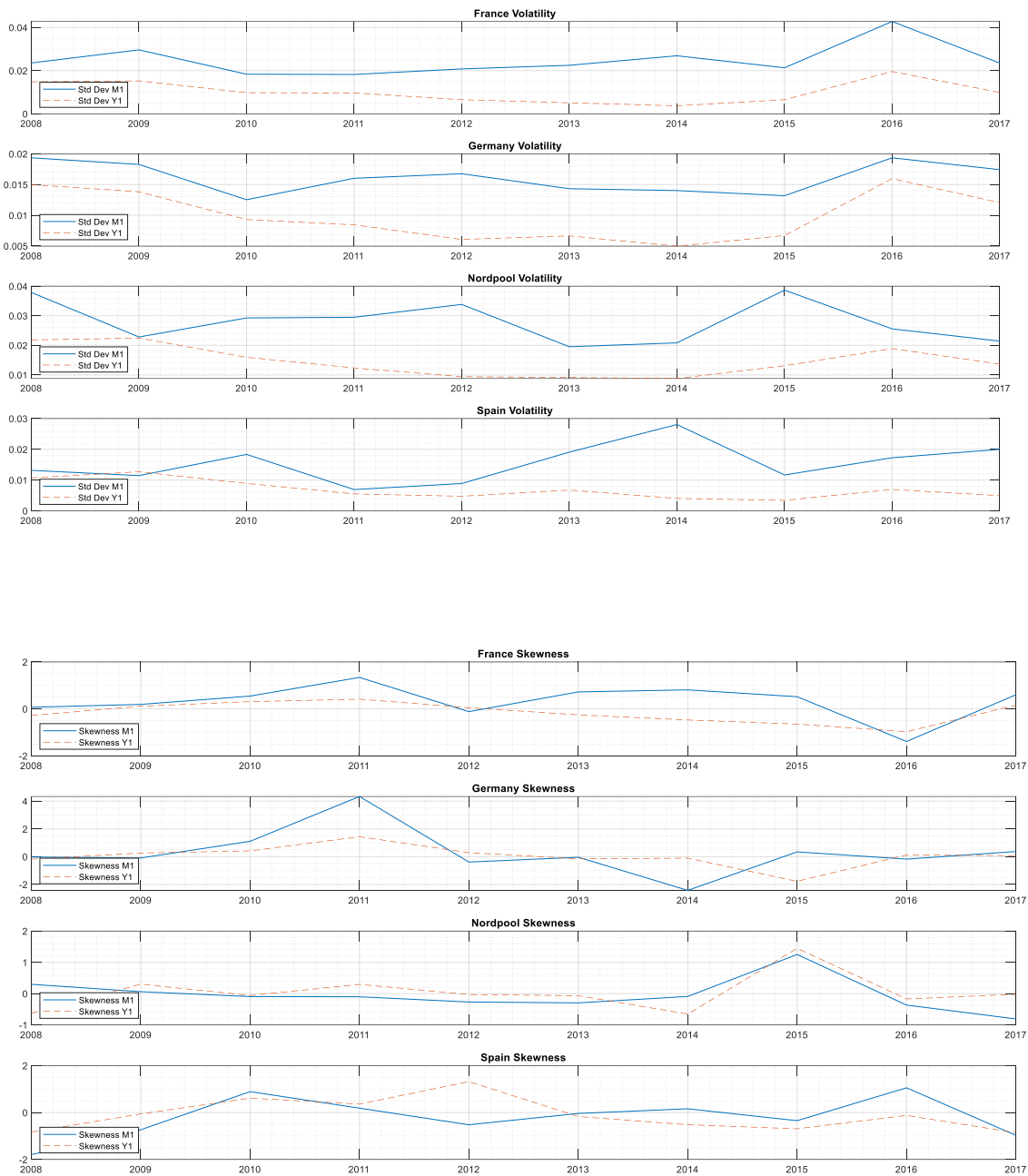

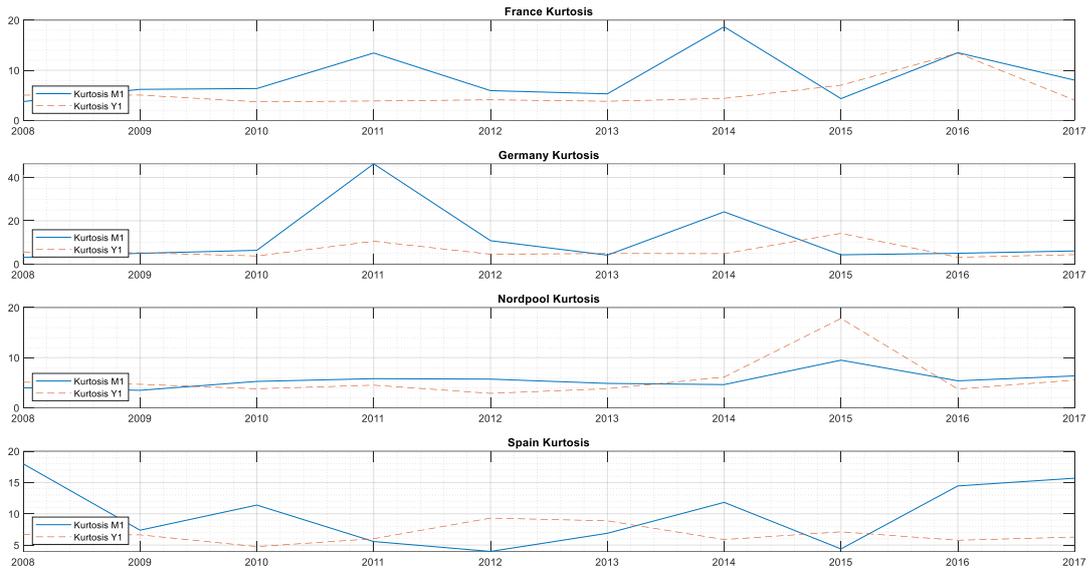

**Figure 4**: VaR Forecasts with all methods. Contract M1, German market.

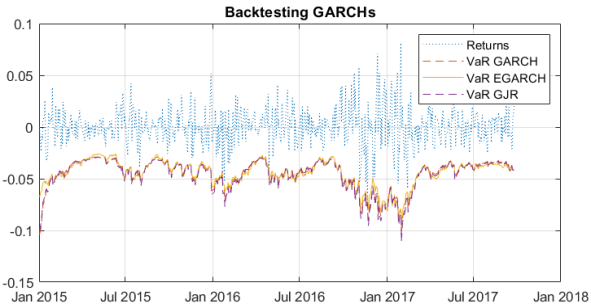
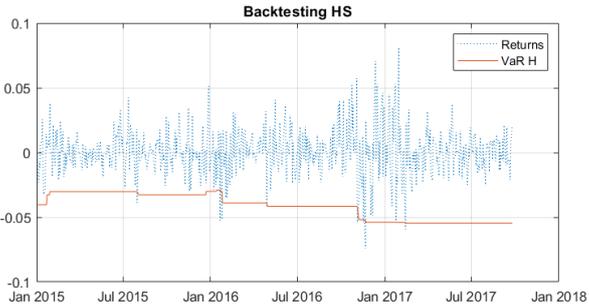
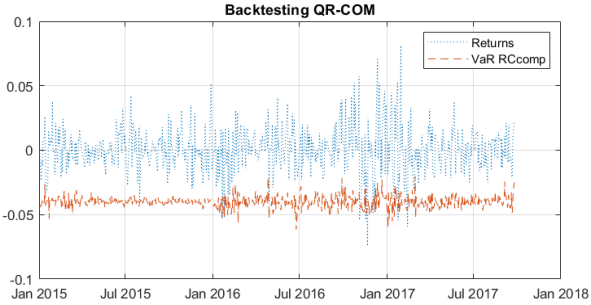
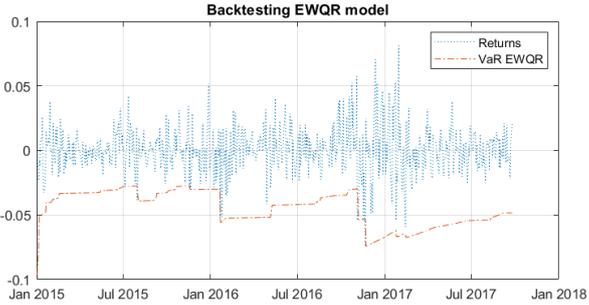